# Scalability Terminology:
# Farms, Clones, Partitions, and Packs:
# RACS and RAPS

Bill Devlin, Jim Gray, Bill Laing, George Spix

# Scalability Terminology:
# Farms, Clones, Partitions, and Packs:
# RACS and RAPS

Bill Devlin, Jim Gray and Bill Laing, George Spix
Microsoft
{BillDev, Gray, Blaing, GSpix}@Microsoft.com, http://research.microsoft.com/~gray
December 1999

Server systems must be able to start small and grow as demand increases. Internet-based eCommerce has made system growth more rapid and dynamic. Application Service Providers that consolidate application processing into large sites must also need dynamic growth. These sites grow by *scale up*, replacing servers with larger servers, or they grow by *scale out* adding extra servers. The scale out approach gives the slogans **buying computing by the slice** and **building systems from CyberBricks**: the brick or slice is the fundamental building block.

The collection of all the servers, applications, and data at a particular site is called a ***farm***. Farms have many functionally specialized ***services***, each with its own applications and data (e.g. directory, security, http, mail, database, etc.,). The whole farm is administered as a unit, having common staff, common management policies, facilities, and networking.

For disaster tolerance, a farm's hardware, applications, and data are duplicated at one or more geographically remote farms. Such a collection of farms is called a ***geoplex***. If a farm fails, the others continue offering service until the failed site is repaired. Geoplexes may be *active-active* where all farms carry some of the load, or *active-passive* where one or more are hot-standbys.

Farms may grow in two ways: (1) *cloning* or (2) *partitioning*. A service can be ***cloned*** on many replica nodes each having the same software and data. Requests are then routed to individual members of the clone set. For example, if a single-node service becomes overloaded, the administrator can duplicate the node's hardware, software, and data on a second node, and then use a load-balancing system to allocate the work between those two nodes. Load balancing can be external to the clones (e.g., an *IP sprayer* like Cisco LocalDirector™), or internal to them (e.g., an *IP sieve* like Network Load Balancing.)[1]
The collection of clones for a particular service is called a ***RACS (Reliable Array of Cloned Services)***. Cloning and RACS have many advantages. Cloning offers both scalability and availability. If one clone fails, the other nodes can continue to offer service, perhaps with degraded performance because they may be overloaded. If the node and application failure detection mechanisms are integrated with the load-balancing system or with the client application, then clone failures can be completely masked. Since clones are identical, it is easy to manage them: administrative operations on one service instance at one node are replicated to all others[2]. As a rule of thumb, a single administrator can manage an appropriately designed service running on hundreds of clones (a RACS of hundreds of nodes).

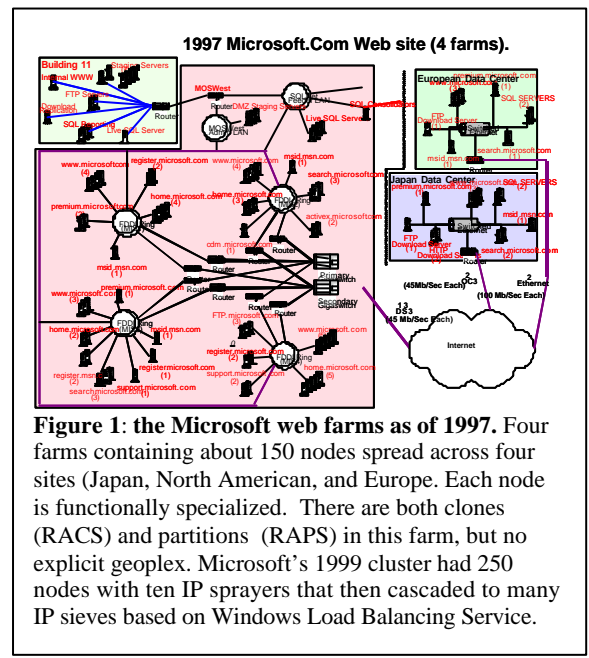

**Figure 1**: **the Microsoft web farms as of 1997.** Four farms containing about 150 nodes spread across four sites (Japan, North American, and Europe. Each node is functionally specialized. There are both clones (RACS) and partitions (RAPS) in this farm, but no explicit geoplex. Microsoft's 1999 cluster had 250 nodes with ten IP sprayers that then cascaded to many IP sieves based on Windows Load Balancing Service.

---

[1] We mix the terms node, process and server, implicitly assuming functionally specialized nodes. It is quite possible for many partitions and clones to run many processes on a single physical server node. A huge SMP server may support many partitions or clones. Using large servers as bricks makes management easier by reducing the number of bricks.

RACS and cloning are an excellent way to add processing power, network bandwidth, and storage bandwidth to a farm. But, ***shared-nothing RACS***, in which each clone duplicates all the storage locally, is not a good way to grow storage capacity. Each clone has identical storage, and all updates must be applied to each clone's storage. So, cloning does not improve storage capacity. Indeed, cloning is problematic for write-intensive services since all clones must perform all writes, so there is no improvement in throughput, and there are substantial challenges in correctly performing the concurrent updates. Clones are best for read-only applications with modest storage requirements.

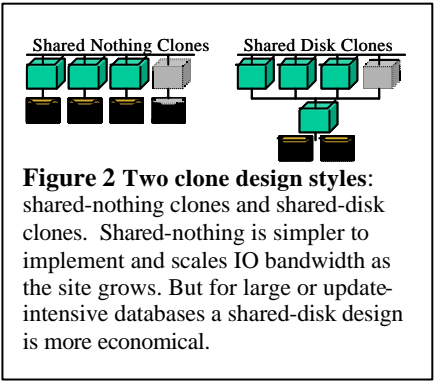

**Figure 2 Two clone design styles**: shared-nothing clones and shared-disk clones. Shared-nothing is simpler to implement and scales IO bandwidth as the site grows. But for large or update-intensive databases a shared-disk design is more economical.

One way to ameliorate the cost and complexity of cloned storage is to let all the clones share a common storage manager. This ***shared-disk RACS*** design, often called a *cluster* (VaxCluster, Sysplex, or Storage Area Network), has stateless servers each accessing a common backend storage server (see Figure 2). This design requires the storage server to be fault-tolerant for availability, and still requires subtle algorithms to manage updates (cache invalidation, lock managers, transaction logs, and the like). As the system scales up, the update traffic can become a performance bottleneck. Despite these shortcomings, shared-disk RACS have many advantages. They have been a popular design for 20 years.

***Partition***s grow a service by duplicating the hardware and software, and by dividing the data among the nodes. In essence it is like the shared-nothing clone of Figure 2, but only the software is cloned, the data is divided among the nodes. Partitioning adds computation power, storage capacity, storage bandwidth, and network bandwidth to the service each time a node is added.

Ideally, when a partition is added, the data is automatically repartitioned among the nodes to balance the storage and computational load. Typically, the application middleware partitions the data and workload by object. For example, mail servers partition by mailboxes, while sales systems might partition by customer accounts or by product lines.

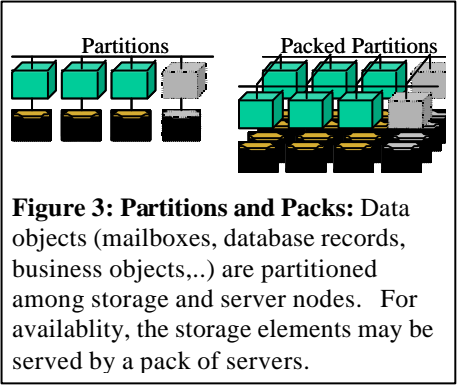

**Figure 3: Partitions and Packs:** Data objects (mailboxes, database records, business objects,..) are partitioned among storage and server nodes. For availablity, the storage elements may be served by a pack of servers.

---

[2] In some designs all the clones have a common boot disk that stores all their software and state, this is called a *shared-disk clone*. In general, clones have identical state except for their physical network names and addresses.

The partitioning should automatically adapt as new data is added and as the load changes.

Partitioning is *transparent* to the application Requests sent to a partitioned service are routed to the partition with the relevant data. If the request involves data from multiple partitions (e.g. transfer funds from one account to another), then the application sees the multiple business objects as though they were all local to that application. Transparency and load balancing are difficult technical tasks, but many systems implement them. Linear scaling is possible if each request accesses only a few partitions. Incremental growth re-partitions the data so that some "buckets" of data move to the new node.

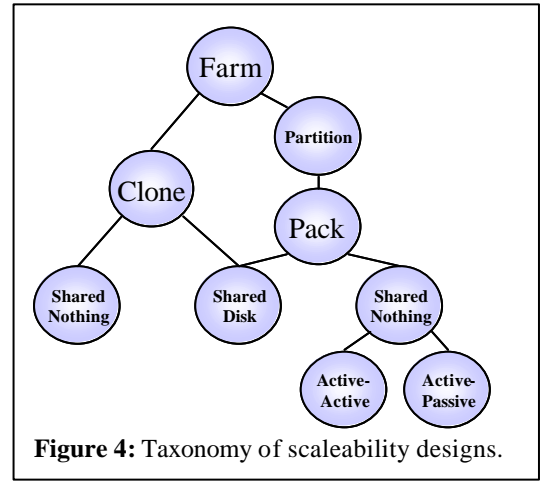

**Figure 4:** Taxonomy of scaleability designs.

Partitioning does not improve availability because the data is stored in only one place. If a disk fails or if the server that manages that disk fails, then that part of the service is unavailable-- that mailbox is not readable, that account cannot be debited, or that patient-record cannot be found. Unlike shared-nothing cloning, which adds redundant storage; simple partitioning (and shared-disk cloning) has just one copy of the data. A *geoplex* guards against this loss of storage, but it is fairly common to locally *duplex* (raid1) or *parity protect* (raid5) the storage so that most failures are masked and repaired.

Even if the storage media is fault-tolerant, a partition might fail due to hardware or software. To give the illusion of instant repair, partitions are usually implemented as a *pack* of two or more nodes that provide access to the storage. These can either be *shared-disk pack* or *shared-nothing packs*. That is, either all members of the pack may access all the disks (a shared-disk partition), or each member of the pack may serve just one partition of the disk pool during normal conditions (a shared-nothing partition), but serve a failed partition if the partition's primary server fails. The shared-disk pack is virtually identical to a shard-disk clone, except that the pack is serving just one part of the total database.

A shared-nothing pack offers two options: each member of the pack can have primary responsibility for one or more partitions. All requests with an affinity to that partition will be routed to that node, and each member of the pack is actively serving some partition. When a node fails, the service of its partition migrates to another node of the pack. This is called the *active-active pack* design. If just one node of the pack is actively serving the requests and the other nodes are acting as hot-standbys, it is called an *active-passive pack*.

By analogy with cloned servers and RACS, the nodes that support a packed-partitioned service are collectively called a **Reliable Array of Partitioned Services (RAPS).** RAPS provide both

| | | | |
|---|---|---|---|
| **Farm** | **Clone** | | **Shared nothing**: rack & stack ISP web servers or RAS servers |
| | | | **Shared disk:** Clusters: VaxCluster, Sysplex, EMC |
| | **Partition** | **Pack** | **Shared disk**: similar to clone shared disk |
| | | | **Shared nothing**: mail and database servers protected with fail over: Tandem, Teradata, Microsoft MSCS, … |

scalability and availability.

Multi-tier applications use both RACS and RAPS (See Figure 5). A hypothetical application consists of a *front tier* that accepts requests and returns formatted responses, a *middle-tier* of stateless business logic and a *data-tier* that manages all writeable state. RACS work well in the front and middle tiers since all the processing is stateless. RAPS are required for the data tier. RACS are easier to build, manage, and incrementally scale. So maximizing the use of RACS is a design goal. Multi-tier application designs provide the functional separation that makes this possible.

Load balancing and routing requirements are different at each tier. At the front tier, IP-level load distribution schemes give reasonable balancing assuming there is a large set of potential clients and requests have no affinity. The middle-tier understands the request semantics, and so can make data and process specific load steering decisions. At the data tier the problem is routing to the correct partition.

## Software Requirements for GeoPlexs, Farms, RACS, and RAPS

The Microsoft website of Figure 1 is daunting: it represents about ten million dollars of equipment, a huge monthly telecommunications bill, and several million dollars worth of buildings. It has over 10 TB of storage, and 3 Gbps of bandwidth to the Internet. But that was 1997, in the last two years, the capacity has increased about three-fold, and the site has nearly three hundred nodes. In addition, a sister farm, HotMail™ has more than two thousand nodes. Both these sites add a few nodes per day. This story is repeated at many other sites around the world: AOL, Yahoo, Amazon, Barnes&Noble, eSchwab, eBay, LLBean, and many others report rapid growth and change in their web sites. Many of these sites are in fact hosted at facilities built with the sole purpose of co-locating multiple large web sites close to redundant high bandwidth Internet connectivity.

This following is more of a wish list than a reflection of current tools and capabilities, but the requirements are fairly easy to state. The first requirement for such a huge site is that it must be possible to manage everything from a single remote console treating RACS and RAPS as entities. Each device and service should generate exception events that can be filtered by an automated operator. The operations software deals with "normal" events, summarizes them, and helps the operator manage exceptional events: tracking the repair process and managing farm growth and evolution. The operations software recognizes the failures and orchestrates repair. This is a challenge when request processing spans multiple functional tiers. Automated operations simplify farm management but are even more important in guaranteeing site availability. Automation reduces manual operations procedures and reduces the chance of operator error. Both the software and hardware components must allow online maintenance and replacement. Tools that support versioned software deployment and staging across a site are needed to manage the upgrade process in a controlled manor. This applies to both the application and system software. Some large Internet sites deploy application modifications weekly or even daily. System software changes are much less frequent but the results of a deployment mistake can be disastrous.

Building a farm requires good tools to design user interfaces, services, and databases. It also requires good tools to configure and then load balance the system as it evolves. There are adequate tools today, and they are making enormous progress over time. It is now fairly easy to build and operate small and medium-sized web sites, but large systems (more than 1M page views per day) are still daunting. Multi-tier application design that enables both RACS and RAPS to be used in combination is still an art and improved design tools could help considerably.

Clones and RACS can be used for read-mostly applications with low consistency requirements, and modest storage requirement (less than 100 GB or about $1,000 today). Web servers, file servers, security servers, and directory servers are good examples of cloneable services. Cloned services need automatic replication of software and data to new clones, automatic request routing to load balance the work, route around failures, and recognize repaired and new nodes. Clones also need simple tools to manage software and hardware changes, detect failures, and manage the repair.

Clones and RACS are not appropriate for stateful applications with high update rates. Using a shared-disk clone can ameliorate some of these problems, but at a certain point the storage server becomes too large and needs to be partitioned. Update-intensive and large database applications are better served by routing requests to servers dedicated to serving a partition of the data (RAPS). This affinity routing gives better data locality and allows caching of the data in main memory without paying high cache-invalidation costs. Email, instant messaging, ERP, and record keeping are good examples of applications that benefit from partitioned data and affinity routing. Each of these applications is nicely partitionable, and each benefits from partitioned scale out. In addition, database systems can benefit from parallel searching, running one query in parallel using many processors operating on many disks. For availability, partitioned systems require some form of packing: so that if one node fails, the stateful service (and its state) can quickly migrate to a second node of the pack.

Partitioned systems need the manageability features of cloned systems, but in addition the middleware must provide transparent partitioning and load balancing. This is an application-level service provided by the mail system (automatically migrate mailboxes to new servers), database systems (spilt and merge data partitions), and other middleware. The middleware software uses the operating system fail-over mechanism (packs) to create a highly available service. The services also expect to program the request routing system to route requests to the appropriate service partition.

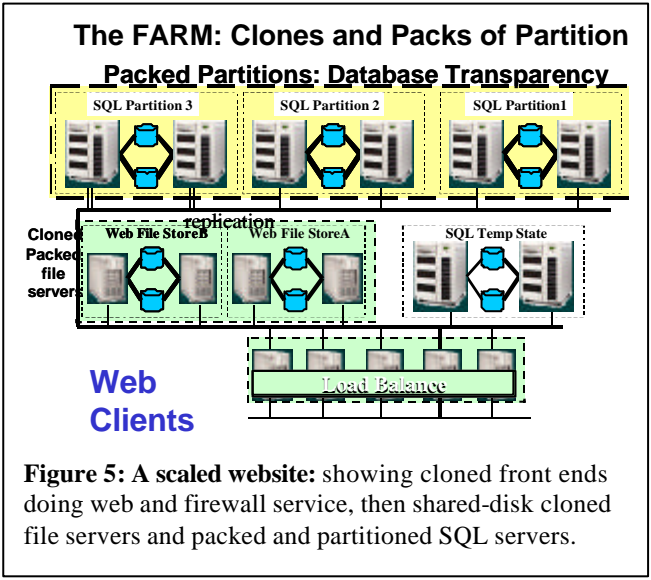

**Figure 5: A scaled website:** showing cloned front ends doing web and firewall service, then shared-disk cloned file servers and packed and partitioned SQL servers.

## Performance and Price/Performance Metrics

Figure 1 represents a huge system. One cannot buy a single 60 billion-instructions per second processor, or a single 100 TB storage server. So some degree of cloning and partitioning is required.

What is the right building block for a site? Will it be an IBM mainframe (OS390) or a Sun UE1000, or will it be Intel-based servers, or will it be the rack-mounted pizza boxes?

This is a hotly debated topic. The mainframe vendors claim that their hardware and software delivers 5-nines of availability (less than 5 minutes outage per year), and that their systems are much easier to manage than cloned PCs. But mainframe prices are fairly high – 3x to 10x more expensive based on TPC results and anecdotal evidence. There is similar controversy about using commodity servers for database storage. We believe that commodity servers and storage are a very good investment, but we know many others who do not.

No matter what, there is clear consensus that a homogenous site (all NT, all FreeBSD, all OS390) is much easier to manage than a site with many hardware and software vendors. So, once you pick your CyberBricks, you will likely stick with them for several generations.

More to the point, middleware like Netscape, IIS, DB2, Oracle, SQL Server, Notes, Exchange, SAP, PeopleSoft, and Baan are where the administrators spend most of their time. Most tasks are per-website, per-mailbox, per user, or per-customer, not per node. Since web and object services are so CPU intensive, it is easy to see why most web sites use inexpensive clones for that part of the service. In addition to this advantage, we believe commodity software is considerably easier to manage than the traditional services that presume very skilled operators and administrators.

## Summary

The key scalability technique is to replicate a service at many nodes. The simplest form of replication, copies both programs and data. These shared-nothing clones can be as easy to manage as a single instance – yet they provide both scalability and availability (RACS).

Shared-nothing clones are not appropriate for large databases or update-intensive services. For these applications, services can be mapped onto packed-partitions. Packs make partitions highly available by automatically restarting a failed partition on another node with access to the failed partition's storage. Middleware is responsible for making the management of these partitions as simple as the management of a single node (RAPS).

To guard against disaster, the entire farm is replicated at a remote site to build a geoplex.

# Glossary

**Active-Active:** A *pack or geoplex* architecture in which all members are actively processing some work (in contrast to *active-passive)*.

**Active-Passive:** A *pack or geoplex* architecture in which one member is actively processing work and the other member(s) is passively waiting for *fail-over* (in contrast to *active-passive)*.

**Availability**: The fraction of the presented requests that a system services within the required response time.

**Clone**: A replica of a server or service. The clones of a service are called a *RACS*. Requests are distributed among the clones within a *RACS*.

**CyberBrick**: The unit of hardware growth in a farm, often it is a commodity system that is added to a RACS or RAPS.

**Fail-over**: A *partition* may fail on one node and be restarted on a second node of a pack, and a *RACS* or *RAPS* may fail on one *farm* of a *geoplex* and be restarted on a second farm of a geoplex.

**Farm**: A site containing many servers and services, but managed as a single administrative entity. A farm contains RACS and RAPS. A farm may be part of a geoplex.

**Geoplex:** A farm that is replicated at two or more sites, so that if one site has a catastrophic failure, the second site can service the load and thereby provide continuous availability.

**Load Balancing:** The process of distributing requests among clones of a RACS and distributing partitions among members of a pack in order to provide better response time.

**Pack:** A collection of servers that can each host a *partition*. When a partition's current server fails, the partition *fails over* to another member of its pack. Packs improve *availability*.

**Partition:** A part of a service that has been divided among a *RAPS*. Each partition services a specific part of the overall service. Mail servers and database servers are often partitioned in this way.

**RACS (Reliable Arrays of Cloned Services):** A collection of clones all performing some service. Requests are directed to the RACS, and processed by one of the clones. The RACS is managed as a single entity.

**RAID (Reliable Array of Independent Disks):** A group of disks that are aggregated to improve availability, bandwidth, or management.

**RAPS (Reliable Arrays of Partitioned Services):** A collection of clones all performing some service. Each request to the RAPS is directed to the appropriate partition and processed by that partition. The RAPS is managed as a single entity.

**Scalability**: The ability to grow the power or capacity of a system by adding components.

**Scale Up**: Expanding a system by incrementally adding more devices to an existing node, typically by adding cpus, disks, and NICs to a node.

**Scale Out**: Expanding a system by adding more nodes, complete with processors, storage, and bandwidth.

**Shared Disk:** A *pack, clone, or geoplex* architecture in which disks and state are shared among the services. In a *packed partitioned* architecture, the disks may *fail-over* when the *partition* migrates to a new member of the pack.

**Shared Nothing:** A *pack, clone, or geoplex* architecture in which disks and state are not shared among the services – rather the state is replicated at each clone or pack member. In a *packed partitioned* architecture, the disks do not *fail-over* when the *partition* migrates to a new member of the pack, rather the partition uses the local replica of the state.

**Transparency:** In general hiding implementation details from the clients. In the context of scalability, hiding the partitioning, cloning, and geoplexing from the clients. Client requests are automatically routed to the correct partition or clone.